\documentstyle[psfig,frontieres,art10]{article} 
\newcommand{\nc}[2]{\newcommand{#1}{#2}}
\newcommand{\ncx}[3]{\newcommand{#1}[#2]{#3}}
\ncx{\pr}{1}{#1^{\prime}}
\nc{\nl}{\newline}
\nc{\np}{\newpage}
\nc{\nit}{\noindent}
\nc{\be}{\begin{equation}}
\nc{\ee}{\end{equation}}
\nc{\ba}{\begin{array}}
\nc{\ea}{\end{array}}
\nc{\dsp}{\displaystyle}
\nc{\fs}{\footnotesize}
\nc{\bit}{\bibitem}
\nc{\ct}{\cite}
\ncx{\dd}{2}{\frac{\partial #1}{\partial #2}}
\nc{\pl}{\partial}
\nc{\nb}{\nabla}
\nc{\dg}{\dagger}
\nc{\ag}{\alpha}
\nc{\bg}{\beta}
\nc{\gam}{\gamma}
\nc{\Gam}{\Gamma}
\nc{\bgm}{\bar{\gam}}
\nc{\del}{\delta}
\nc{\Del}{\Delta}
\nc{\eps}{\epsilon}
\nc{\ve}{\varepsilon}
\nc{\zg}{\zeta}
\nc{\th}{\theta}
\nc{\vt}{\vartheta}
\nc{\Th}{\Theta}
\nc{\kg}{\kappa}
\nc{\lb}{\lambda}
\nc{\Lb}{\Lambda}
\nc{\ps}{\psi}
\nc{\Ps}{\Psi}
\nc{\sg}{\sigma}
\nc{\spr}{\pr{\sg}}
\nc{\Sg}{\Sigma}
\nc{\rg}{\rho}
\nc{\fg}{\phi}
\nc{\Fg}{\Phi}
\nc{\vf}{\varphi}
\nc{\og}{\omega}
\nc{\Og}{\Omega}
\nc{\pog}{\og^{\prime}}
\nc{\px}{x^{\prime}}
\nc{\pnb}{\nabla^{\prime}}
\nc{\ha}{\hat{a}}
\nc{\hb}{\bar{h}}
\nc{\hh}{\hat{h}}
\nc{\htt}{\hat{t}}
\nc{\hR}{\hat{R}}
\nc{\bzg}{\bar{\zg}}
\nc{\sh}{\tilde{h}}
\nc{\sT}{\tilde{T}}
\nc{\sth}{\tilde{\th}}
\nc{\srg}{\tilde{\rho}}
\nc{\sPi}{\tilde{\Pi}}
\nc{\cI}{{\cal I}}
\nc{\cL}{{\cal L}}
\nc{\cH}{{\cal H}}
\nc{\cE}{{\cal E}}
\nc{\uh}{\underline{h}}
\nc{\up}{\underline{\pi}}
\nc{\uT}{\underline{T}}
\nc{\bv}{\mbox{\boldmath $v$}} 
\nc{\bor}{\mbox{\boldmath $r$}} 
\nc{\bR}{\mbox{\boldmath $R$}} 
\nc{\bL}{\mbox{\boldmath $L$}} 
\nc{\vq}{\mbox{$\vec{q}$}}
\nc{\qp}{\mbox{$\pr{\vec{q}\,}$}}
\nc{\vp}{\mbox{$\vec{p}$}}
\nc{\va}{\mbox{$\vec{a}$}}
\nc{\vb}{\mbox{$\vec{b}$}}
\nc{\Ztwo}{\mbox{\sf Z}_{2}}
\nc{\Tr}{\mbox{Tr}}
\nc{\lh}{\left(}
\nc{\rh}{\right)}
\ncx{\Gm}{3}{ \Gamma_{{#1}{#2}}^{\:\:\:\:\:{#3}} }
\ncx{\vs}{1}{\vspace{#1 ex}}

\begin{document} 

\heading{GRAIL \\
an omni-directional gravitational-wave detector}

\par\medskip\noindent 
\author{D.\ van Albada$^{\fs a}$, W.\ van Amersfoort$^{\fs d}$, 
H.\ Boer Rookhuizen$^{\fs d}$, J.\ Flokstra$^{\fs e}$, \nl  
G.\ Frossati$^{\fs c}$,  H.\ van der Graaf$^{\fs d}$, 
A.\ Heijboer$^{\fs d}$,  G.\ Heijboer$^{\fs d}$, \nl 
E.\ van den Heuvel$^{\fs a}$, J.W.\ van Holten$^{\fs d}$,  
G.J.\ Nooren$^{\fs d}$, J.\ Oberski$^{\fs d}$, \nl 
H.\ Rogalla$^{\fs e}$, A.\ de Waele$^{\fs b}$, 
P.\ de Witt Huberts$^{\fs d}$ 
}
\nit
\footnoterule 
\nit {\fs 
a.\ University of Amsterdam,
b.\ Technical University, Eindhoven,  
c.\ University of Leiden, \nl 
d.\ NIKHEF, Amsterdam,    
e.\ University of Twente }

%

 
\begin{abstract} 
An cryogenic spherical and omni-directional resonant-mass detector proposed 
by the GRAIL collaboration is described. 
\end{abstract} 
\vs{1}
\nit
\footnoterule 
\vs{2}
 
\nit
Spherical resonant-mass detectors have been discussed in the 
literature since the early 70's \ct{Forw,Pk1,W}. They have a number 
of obvious advantages over Weber bars: 
\begin{itemize}
\item Spherical detectors have the largest mass for a given linear 
      dimension. 
\item They are always optimally oriented with respect to any source. 
\item They can determine the transverse plane of polarization 
      of a signal, allowing the reconstruction of the direction 
      of the source (modulo a reflection in the transverse plane).  
\item They are sensitive to a scalar component of gravitational 
      radiation. 
\end{itemize} 
Moreover, in view of recent developments in the design of 
electro-mechanical transducers the fractional bandwidth of a 
spherical detector, equipped with a sufficient number of such 
transducers, is expected to reach values of the order of 20\%, an 
improvement by two orders of magnitude over resonant detectors 
presently in operation. The development of a spherical resonant-mass 
detector is therefore considered an attractive new step towards the 
goal of observing gravitational radiation from astrophysical sources
\ct{ZhM,JM,CL}. 
\vs{1}

\nit 
Recently we presented a preliminary design for a 115 ton spherical 
detector, to be made out of a high-Q alloy such as Cu-Al ($Q \geq 
10^7$), and operated at a temperature in the range 10-20 mK 
\ct{GR}. This is to be achieved by integrating a high-power 
$^3$He-$^4$He dilution refrigerator into the system. 

The free vibration modes of a perfect sphere can be computed 
analytically \ct{Love,AD,Pk1,Lobo1}. In general the 
displacement field $\vec{u}(\bor,t)$ can be expanded as 

\be
\vec{u}(\bor,t)\, =\, \sum_k\, a_k(t) \vec{\psi}_k(\bor),
\label{1.1}
\ee

\let\picnaturalsize=N
\def\picsize{2.0in}
\def\picfilename{gwfig1.eps}
\ifx\nopictures Y\else{\ifx\epsfloaded Y\else\input epsf \fi
\let\epsfloaded=Y
\centerline{\ifx\picnaturalsize N\epsfxsize \picsize\fi \epsfbox{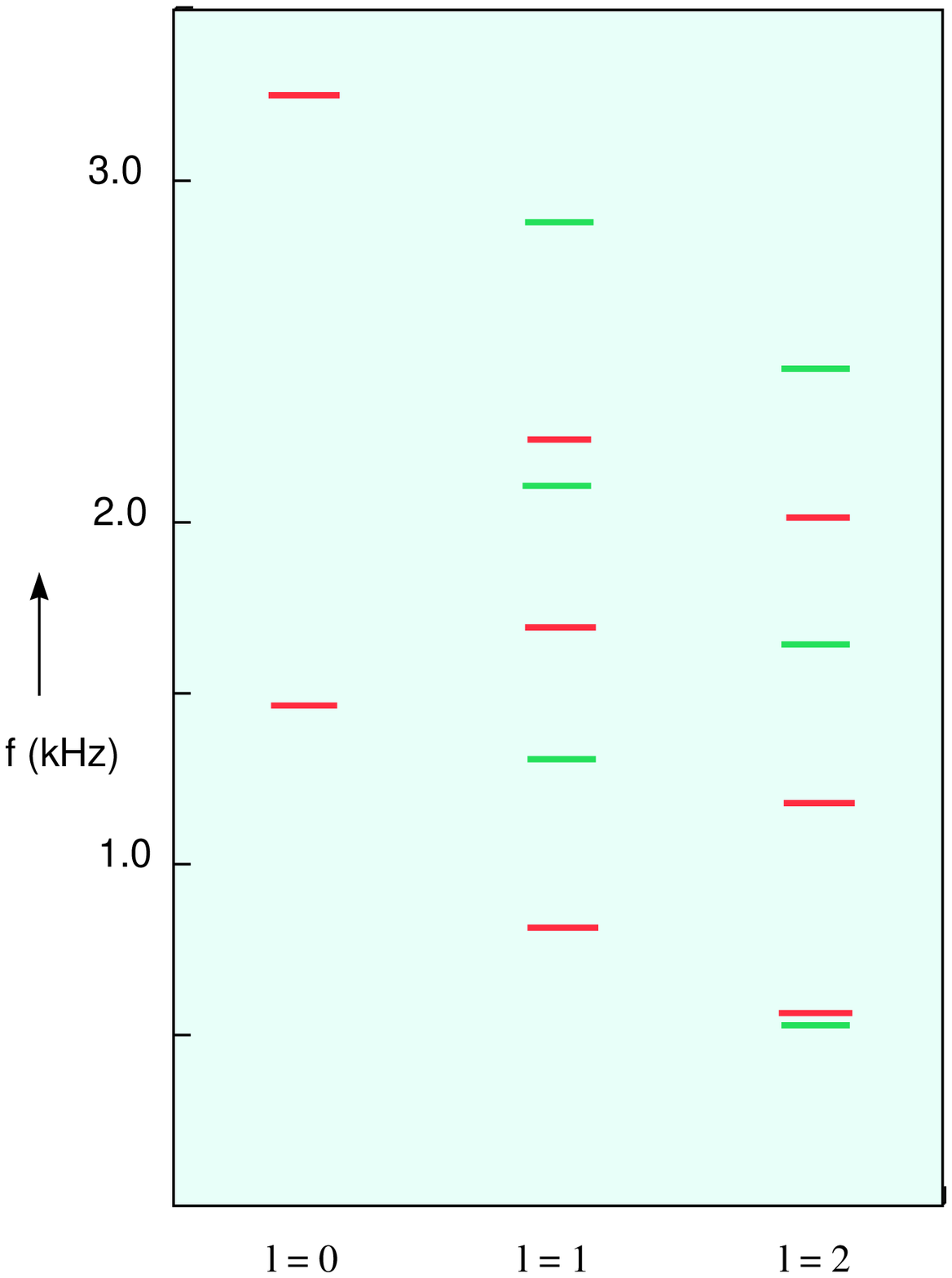}}}\fi
\begin{center} 
Fig.1 Red: spheroidal mode; green: toroidal modes 
\end{center} 
\vs{1} 

\nit
where $k$ collectively denotes the characteristic numbers $(n,l,m)$ 
of the free vibrations, with $n$ the number of zero's of the radial 
component and $(l,m)$ specifying the angular dependence in terms of the 
$2^l$-multipole character. In the absence of dissipation, the time 
dependence of the modes is described by the oscillator equation 

\be
\ddot{a}_k\, +\, \og_k^2 a_k\, =\, 0, 
\label{1.2}
\ee

\nit
with $\og_k$ the angular frequency of the free vibration mode. 
Depending on the reflection parity, modes may be classified as 
spheroidal or toroidal; only the spheroidal modes couple to 
gravitational waves. The frequencies $f_k = \og_k/2\pi$ 
are plotted for the lowest spheroidal and toroidal modes of 
various multipole character in fig.1. Clearly, the fundamental mode, 
which for the GRAIL sphere has a frequency close to 700 Hz, is of 
quadrupole type, thus matching the tensor character of the 
gravitational field. 

In the presence of external forces, eq.(\ref{1.2}) for the amplitudes 
is replaced by the inhomogenous equation 

\be 
\ddot{a}_k\, +\, \og_k^2 a_k\, =\, f_k(t),
\label{1.3}
\ee

\nit 
where the driving term $f_k(t)$ is given in terms of the force 
density $\vec{f}(\bor,t)$ by 

\be
f_k(t)\, =\, \frac{1}{M}\, \int d^3r\, \vec{\psi}_k \cdot \vec{f},
\label{1.4}
\ee

\nit 
with $M$ the total mass of the sphere. A weak gravitational wave is 
described by a force density of quadrupole type, i.e.\ $l = 2$; to 
the amplitude $a_m$ of the $m$th quadrupole mode it imparts an 
acceleration 

\be
f_m(t)\, =\,  \frac{1}{2}\, \ddot{h}_m(t) \chi R, \hspace{3em} 
 m = -2, -1, ..., +2. 
\label{1.5}
\ee

\nit 
The dimensionless constant of proportionality $\chi$ depends 
weakly on the Poisson ratio $\sg$ of the material; its value  
$\chi \approx 0.6$ varies only by a few percent over a large range 
of values of $\sg$. For the GRAIL detector the radius $R = 1.5$ m. 
\vs{1}

\nit 
For example, an exponentially decaying burst of gravitational 
radiation with the time variation of its amplitude of the form 

\be
\dot{h}_m(t)\, =\, \frac{h_m}{\tau}\, \th(t) e^{-t/\tau}, 
\label{1.6}
\ee

\nit
produces for times $t \gg 2 \tau$ an amplitude 

\be
a_m(t)\, =\, a^{(0)}_m(t)\, +\, \frac{h_m}{1 + \og_0^2 \tau^2}\, 
 \lh \cos \og_0 t + \og_0 \tau \sin \og_0 t\rh. 
\label{1.7}
\ee

\nit
Here $\og_0$ is the frequency of the free quadrupole mode, and 
$a^{(0)}_m(t)$ is an arbitrary free vibration of the system, 
providing a background against which the signal is to be measured. 
If the source is at a large distance $d$, and converts an energy 
$E_{burst} = \eta M_{\odot}c^2$ into an isotropic exponential 
burst of gravitational radiation with amplitude decay time $\tau$, 
then a simple calculation shows that typically 

\be
h_m\, =\, \frac{2}{d}\, \sqrt{\eta R_{\odot} c \tau}\, =\, 
          0.6\, \times 10^{-13}\, 
          \frac{\sqrt{\eta \tau}}{\lh d/\mbox{1 kpc}\rh}. 
\label{1.8}
\ee

\nit
with $R_{\odot} \approx 3$ km the Schwarzschild radius of the sun.  
For a 1 ms burst at $d = 10$ Mpc converting 0.1 percent of a solar 
mass into gravitational waves the contribution to the amplitude 
$a_m$ is of the order $h_m/\og_0 \tau \approx 10^{-21}$. 

Under realistic conditions, the detected signal will be accompanied 
by external background (e.g., seismic, electromagnetic, cosmics) 
and internal noise. The GRAIL design incorporates a chain of 
masses and rods to attenuate (by reflection) external vibrations by 
more than 300 dB. This should be sufficient to eliminate high-frequency 
seismic background. A point of special concern will be to avoid 
short-circuiting this attenuation system by the dilution refrigerator 
cooling the sphere via the last masses in the chain. Finally, cosmics 
can be largely eliminated by placing the detector underground. 

Important sources of noise in the detector are the thermal noise of 
the sphere, and the displacement and force noise of the amplifiers. 
As illustrated in fig.2, the latter are conventionally described by 
a noise temperature $T_n$ and a noise impedance $r_n$, defined by:

\be 
kT_n = \sqrt{S_u S_f}, \hspace{3em} r_n = \sqrt{S_f/S_n}. 
\label{1.9}
\ee

\let\picnaturalsize=N
\def\picsize{2.5in}
\def\picfilename{gwfig2.eps}
\ifx\nopictures Y\else{\ifx\epsfloaded Y\else\input epsf \fi
\let\epsfloaded=Y
\centerline{\ifx\picnaturalsize N\epsfxsize \picsize\fi \epsfbox{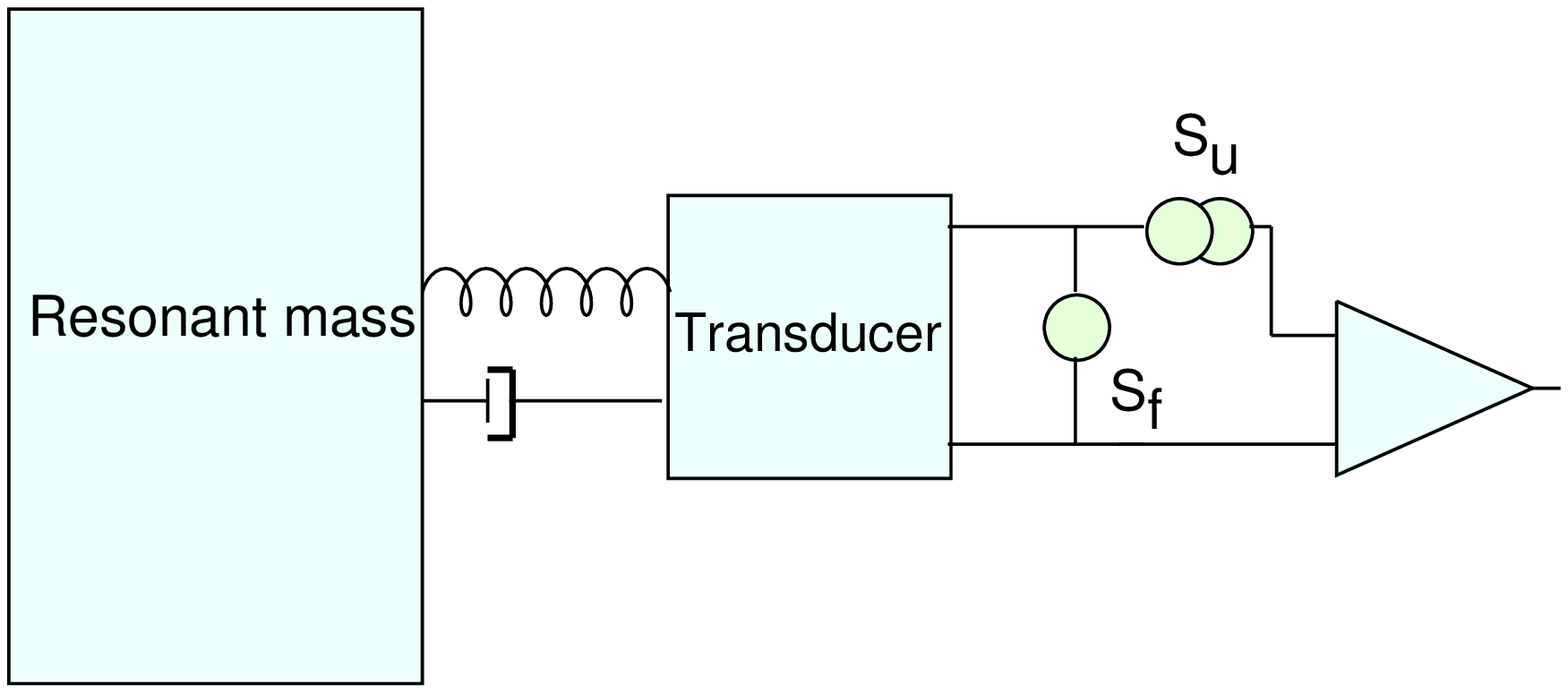}}}\fi
\begin{center} 
Fig.2 Noise model of resonant-mass detector with transducer 
\end{center}
\vs{1} 

\nit 
To optimize the signal-to-noise ratio, one should match the noise 
impedance with the output impedance of the tranducer, and the noise 
temperature with the effective temperature of the sphere, 
$T_{eff} = T/\bg Q$, with $\bg$ the electromechanical coupling 
between the sphere and the amplifier \ct{Piz}. In the optimal case 
the detector is quantum-limited, with $kT_n = h f_0 \approx 0.4 \times 
10^{-30}$ J. To achieve this, a high $Q$-value is indispensible: 
$\bg Q > 10^6$, whilst the amplifier also is to approach the quantum 
limit.  

Using the methods described in \ct{JM,St}, we have computed the 
sensitivity of the GRAIL detector equipped with six double mode 
transducers in a TIGA-configuration. A typical result for the noise 
of a quantum-limited detector, for three values of the noise 
impedance: $r_n = (100; 1000; 10,000)$ N.s/m refered back to the 
equivalent gravitational wave input, is sketched in fig.\ 3. For the 
optimal value $r_n = 1000$ the detector has a noise less than the 
equivalent strain of $10^{-23}/\sqrt{\mbox{Hz}}$ over a range of 
almost 100 Hz in frequency. 
\vs{1}

\let\picnaturalsize=N
\def\picsize{3.0in}
\def\picfilename{gwfig3.eps}
\ifx\nopictures Y\else{\ifx\epsfloaded Y\else\input epsf \fi
\let\epsfloaded=Y
\centerline{\ifx\picnaturalsize N\epsfxsize \picsize\fi \epsfbox{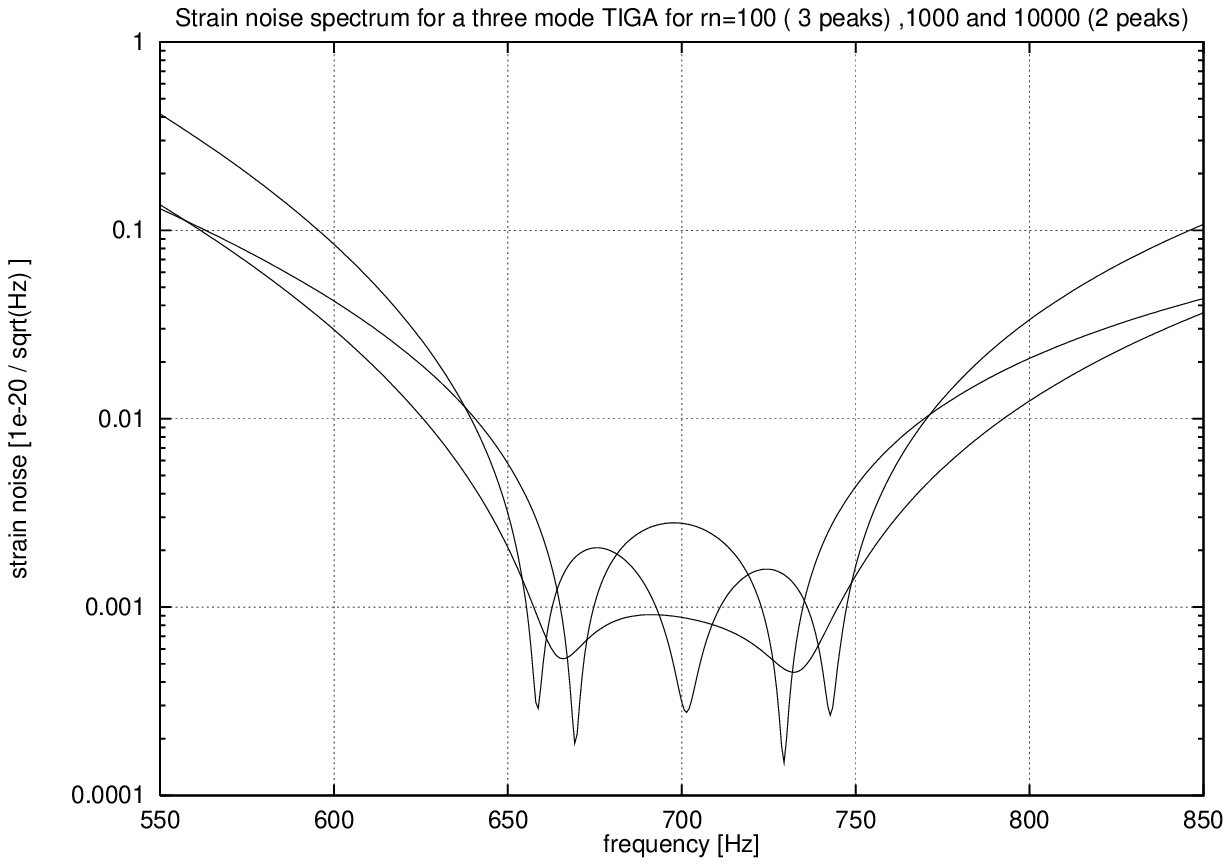}}}\fi
\begin{center}
Fig.3 Sensitivity of quantum limited GRAIL sphere with double-mode 
      transducers 
\end{center}
\vs{1} 
 
\nit 
The sensitivity of a quantum-limited GRAIL is compared to that 
expected for the LIGO-interferometer in its advanced phase \ct{ligo} 
in fig.\ 4. From this figure it may be infered, that the sensitivities  
of the two types of detectors are comparable. In contrast, in  
characteristics in frequency range, and in band width vs.\ 
directional sensitivity, they will be largely complementary. As 
such, the GRAIL detector might become an important element in a 
worldwide network of gravitational wave detectors. 

\let\picnaturalsize=N
\def\picsize{2.3in}
\def\picfilename{gwfig4.eps}
\ifx\nopictures Y\else{\ifx\epsfloaded Y\else\input epsf \fi
\let\epsfloaded=Y
\centerline{\ifx\picnaturalsize N\epsfxsize \picsize\fi \epsfbox{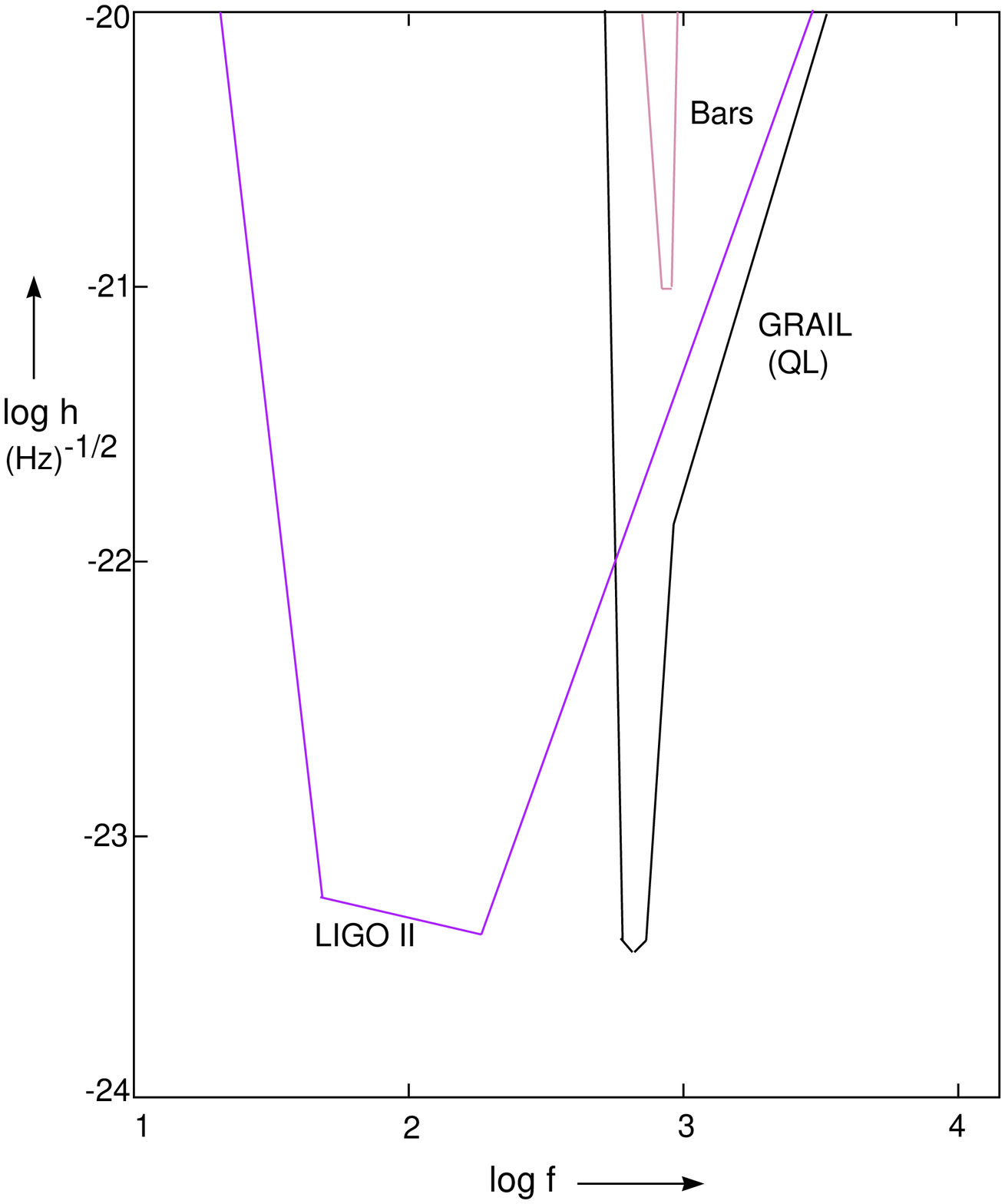}}}\fi
\begin{center} 
Fig. 4 Projected sensitivity of quantum-limited 
GRAIL sphere \\ 
compared to advanced LIGO and cryogenic Weber bars 
\end{center} 

%
 
 
\begin{iapbib}{99} 
\bit{Forw} R.\ Forward, Gen.\ Rel.\ and Grav.\ 2 (1971), 149 
\bit{Pk1} R.V.\ Wagoner and H.J.\ Paik, in: {\em Gravitatione 
          Sperimentale,} Ac.\ Naz.\ dei Lincei (Rome, 1977) 
\bit{W} S.\ Weinberg, {\em Gravitation and Cosmology} 
       (J.\ Wiley, N.Y., 1972), ch.\ 10  
\bit{ZhM} C.\ Zhou and P.\ Michelson, Phys.\ Rev.\ D51 (1995), 2517  
\bit{JM} W.\ Johnson and S.\ Merkowitz, Phys.\ Rev.\ D52 (1995), 2546  
\bit{CL} E.\ Coccia, J.A.\ Lobo and J.A.\ Ortega, Phys.\ Rev.\ D52
         (1995), 3735 
\bit{GR} D.\ van Albada et al., {\em GRAIL, an Omnidirectional 
         Gravitational Wave Antenna}, proposal to the Netherlands 
         Organization for Scientific Research (NWO) (May, 1997) 
\bit{Love} A.E.H.\ Love, {\em A treatise on the mathematical theory 
           of elasticity} (Cambridge Univ.\ Press, 1927) 
\bit{AD} N.\ Ashby and J.\ Dreitlein, Phys.\ Rev.\ D12 (1975), 336 
\bit{Lobo1} J.A.\ Lobo, Phys.\ Rev. D52 (1995), 591 
\bit{Piz} G.\ Pizzella, in: {\em The detection of gravitational 
          waves}, ed.\ D.\ Blair (Cambridge Univ.\ Press, 1991) 
\bit{St} T.R.\ Stevenson, Phys.\ Rev.\ D56 (1997), 564  
\bit{ligo} K.S.\ Thorne, {\em Gravitational radiation}, 17th Texas 
           Symposium on Relativistic Astrophysics and Cosmology;
           Ann.\ N.Y.\ Ac.\ Sc.\ vol. 759 (1995), 127   
\end{iapbib} 
\vfill
\end{document}